\documentclass{elsart}

\IfFileExists{srcltx.sty}{\usepackage[active]{srcltx}}%

\usepackage{amssymb}

\usepackage{graphicx}% Include figure files
\usepackage{dcolumn}% Align table columns on decimal point
\usepackage{bm}% bold math
\usepackage{amsmath}
\begin{document}

\newcommand{\be}{\begin{equation}}
\newcommand{\ee}{\end{equation}}
\newcommand{\ba}{\begin{eqnarray}}
\newcommand{\ea}{\end{eqnarray}}
\def\bone{$B^{(1)}$}
\def\bone{B^{(1)}}
\def\etal{{\it et al.~}}
\def\eg{{\it e.g.~}}
\def\ie{{\it i.e.~}}
\def\DM{dark matter~}
\def\DE{dark energy~} 
\def\GC{Galactic center~} 
\def\susy{SUSY~}
\begin{frontmatter}

\title{Gamma ray bursts and the origin of galactic positrons}
\author{Gianfranco Bertone$^a$, Alexander Kusenko$^b$}
\author{Sergio Palomares-Ruiz$^{b,c}$, Silvia Pascoli$^b$, Dmitry
  Semikoz$^{b,d}$} 
\address{$^a$ NASA/Fermilab Theoretical Astrophysics Group, 60510
Batavia IL\\ $^b$ Dept. of Physics and Astronomy, UCLA, Los Angeles,
CA 90095-1547, USA \\  $^c$ Dept. of Physics \& Astronomy, Vanderbilt 
University, Nashville, TN 37235, USA \\ $^d$INR RAS, 60th October 
Anniversary prospect 7a, 117312 Moscow, Russia }
\begin{abstract}
A recent observation of the 511 keV electron-positron annihilation line
from the Galactic bulge has prompted a debate on the origin of the
galactic positrons responsible for this emission. Assuming equilibrium
between annihilation and injection of positrons in the Galaxy, we
investigate the possibility that positrons were produced by past gamma
ray bursts (GRBs). We compare the positron annihilation rate inferred
by the observed 511 keV line, with the predicted injection rate due to 
e$^+$e$^-$ pairs leaking out of individual GRBs and those pair produced
by GRB photons in the dense Molecular clouds at the Galactic
center. We conclude that the proposed scenario is marginally
consistent with observations, and can reproduce the observed
normalization of the annihilation line only for higher-than-average 
values of the GRB rate in the Galaxy and/or of the Molecular clouds 
optical depth for pair production. 
\end{abstract}
\maketitle
\end{frontmatter}

{\bf Introduction --} The observations of the 511~keV emission line
from the Galactic central region with the SPI camera aboard the
INTEGRAL satellite
~\cite{Knodlseder:2003sv,Jean:2003ci,Teegarden:2004ct,Churazov:2004as,Knodlseder:2005yq,Jean:2005af}
have confirmed earlier estimates of the total
flux~\cite{johnson72,johnson73,haymes75,leventhal78,leventhal80,albernhe81,Teegarden:1996fi,purcell97},
and provided some crucial information on the morphology of the
annihilation region, suggesting an azimuthally symmetric Galactic
bulge component with full width at half maximum $\sim8$ degrees, and
with a 2 $\sigma$ uncertainty of $\sim1$
degree~\cite{Knodlseder:2005yq}. The 511 keV line flux in the bulge 
component amounts to $(1.07 \pm 0.03) 10^{-3}$ photons cm$^{-2}$
s$^{-1}$~\cite{Jean:2005af}, a flux that can be explained  in terms of
electron--positron annihilations via positronium occuring at a rate 
\begin{equation} 
\Gamma_{(e^+e^-\rightarrow \gamma\gamma)} \sim  
 10^{50}\, {\rm yr}^{-1} . 
\label{annihilation_rate}
\end{equation} 
This annihilation rate provides strong evidence for a large population
of Galactic positrons, which are difficult to account for with
standard astrophysical sources. To explain their origin, numerous
scenarios have been proposed (see \eg
Refs.~\cite{Knodlseder:2005yq,Dermer:2001wc,Casse:2003fh}), including
production by black holes and pulsars~\cite{dermer88},
microquasars~\cite{Li:1995ev}, radioactive nuclei from past
supernovae, novae, red giants or Wolf-Rayet
stars~\cite{Casse:2003fh,signore88}, a single recent gamma-ray burst 
event~\cite{Furlanetto:2002sb}, cosmic ray interactions with the
interstellar medium~\cite{kozlowsky87} and stellar
flares~\cite{meynet97}. More recently, new scenarios have been
proposed invoking annihilation of light dark
matter~\cite{Boehm:2003bt,Gunion:2005rw}, decay of
axinos~\cite{Hooper:2004qf}, sterile
neutrinos~\cite{Picciotto:2004rp}, scalars with gravitational strength
interactions~\cite{Picciotto:2004rp}, or mirror
matter~\cite{Foot:2004kd}, color superconducting dark
matter~\cite{Oaknin:2004mn}, superconducting cosmic
strings~\cite{Ferrer:2005xv}, moduli decay~\cite{Kawasaki:2005xj},
Q-balls~\cite{Kasuya:2005ay}, pulsar winds~\cite{Wang:2005cq},
primordial and accreting small-mass black
holes~\cite{Frampton:2005fk,Titarchuk:2005uw}.

Some of these scenarios are problematic, while others have a wide
range of uncertainties. For example, the predicted yields and
distributions of positrons from radionuclei synthesized in supernovae
are only marginally compatible with
observations~\cite{Milne:2001zs}. Supernovae produce positrons mainly 
from the decay $^{56}Ni \rightarrow ^{56}Co \rightarrow ^{56}Fe$,
although the fraction of positrons that would escape the SN ejecta is
poorly known. Different estimates suggest that typically 30\%-- 50\%
of galactic positrons may be explained by SNe Ia and massive stars
(SNII/Ib and WR stars).  This, combined with the average SN Ia rate in
the Galaxy, implies that the contribution of massive stars to the
galactic positrons is subdominant (see Ref.~\cite{Milne:2001zs} and
references therein), although a  new estimate of the SN Ia rate and an
appropriate choice of the Galactic magnetic field may reconcile this
scenario with observations~\cite{Prantzos:2005pz}. In
Ref.~\cite{Casse:2003fh}, a new class of SN, SN Ic, interpreted as the 
result of a bipolar Wolf-Rayet explosion, was proposed as a source of
galactic positrons, from the decays of $^{56} Co$. The supernovae of
this class are short and bright and may be associated to
GRBs~\cite{sn2003}, but their rate is unfortunately unknown.  

It was also suggested that the positrons may be produced from
annihilations of relatively light dark matter particles, with mass in
the 1--100~MeV range~\cite{Boehm:2003bt}. It is essential for this
scenario that the dark matter particles are light. The usual dark
matter candidates, with mass in the (0.1--1)~TeV range, are expected
to produce a large number of electrons and positrons through
annihilations. However, they would also produce high-energy gamma
rays. If one requires the electron-positron annihilations to occur at
a high enough rate to explain the 511~keV emission, the associated
flux of gamma rays would exceed the flux observed by EGRET in the
direction of the Galactic Center by several orders of magnitude.
However, if dark matter is made up of particles with masses below the
muon mass, the annihilation produces only electron-positron
pairs~\cite{Boehm:2003bt}.

It was  shown by Beacom et al.~\cite{Beacom:2004pe} that even
if the dark matter particles annihilate only into electron-positron
pairs, the internal bremsstrahlung emission, inevitably associated
with this process, would exceed the Galactic diffuse gamma-ray
emission, unless the dark matter particle is lighter than $\sim 20$
MeV. A similar constraint applies to other proposed sources of
positrons, including the decay of exotic particles, such as
axinos~\cite{Hooper:2004qf}, sterile
neutrinos~\cite{Picciotto:2004rp}, scalars with gravitational strength 
interactions~\cite{Picciotto:2004rp}, and mirror
matter~\cite{Foot:2004kd}. Recently, the upper limit on the 
injection energy of the positrons in the bulge, thus on the
mass of MeV Dark Matter, was brought down to $\approx 3 MeV$ 
through the analysis of in-flight annihilations~\cite{Beacom:2005qv}. 

Here we investigate the possibility that galactic positrons were
produced by gamma-ray bursts (GRB) which took place in our galaxy
(see e.g. Refs.~\cite{Piran:1999kx,Zhang:2003uk} for review of
GRB). This scenario is appealing for several reasons. First, GRB are
known to exist. Second, jets in GRBs are expected to lose energy
through pair production. Positron production in GRBs has been modeled
and was suggested as a way to identify the location of a single recent
GRB~\cite{Furlanetto:2002sb}, due to pair annihilation at the end of
the radiative phase. Here we are instead interested in those positrons
that escape from the fireball and propagate in the bulge, as well as
in those which are pair produced by GRB photons in the field nuclei of
molecular clouds. We stress then that our scenario is substantially 
different from those focusing positrons from heavy nuclei decay in 
hypernova ejecta (see Ref.~\cite{Casse:2003fh}), despite the possible
association between these two class of objects.

{\bf Propagation of Positrons --} Before deriving the rate at which
GRBs inject positrons in the Galaxy, we first study energy losses and
propagation distance of positrons in the Galactic magnetic and
radiation field. At the energies we are interested in, the positrons
propagating into the  intergalactic medium suffer energy losses mainly
via ionization, on a time scale~\cite{Longair}
\begin{equation}
\tau_{\rm ioniz} \sim  10^{7} 
\frac{\gamma}{\log \gamma + 6.2} 
\left( \frac{N_H}{10^5 {\rm m}^{-3}} \right)^{-1} {\rm yrs},
\label{tionization}
\end{equation}
where $N_H$ is the number density of atoms and $\gamma=E/m_e$ is the initial
gamma factor of the positrons with energy $E$. In the Galactic bulge, $N_H \sim
10^5 {\rm m}^{-3}$~\cite{Longair} and the resulting stopping time is $
\tau_{\rm ioniz} \sim  10^7 - 10^8 \, {\rm yrs}$, for positrons with $\gamma 
\simeq 10 - 10^2$. This time scale is much longer than the typical interval
between GRBs in our galaxy (see eq.~(\ref{rate_per_galaxy}) below);
therefore, one can treat the injection as approximately continuous.

During the time $\tau_{\rm ioniz}$, positrons diffuse in the Galactic
magnetic field. Unfortunately, little is known about the properties of
the magnetic field in the bulge. One expects this field to have both
regular and turbulent components~\cite{beck2000}.  There is compelling
evidence of turbulence in the local Galactic magnetic field, where the
largest scale of the turbulent component is $l_{\rm cell}\sim
50\,$pc~\cite{rand89}. If this turbulence is present in the bulge,
then positrons with $\gamma \sim 10^2$ or less, have Larmor radii much
smaller than the characteristic size of turbulence cells. In this
regime, we can write the diffusion coefficient for an electron with energy
$E$ as (see e.g. Ref.~\cite{Blasi:1998xp} and references therein) 
\begin{equation}
D(E) = 3 \times 10^{26} \, \left ( \frac{E}{m_e} \right)^{1/3}
\left( \frac{B}{1 \mu{\rm G}} \right)^{-1/3}
\left( \frac{l_{\rm cell}}{50 {\rm pc}}  \right)^{2/3} {\rm
  cm}^2 {\rm s}^{-1} .
\end{equation} 
This is a phenomenological formula, which only applies when the
turbulent component is comparable with, or smaller than, the regular
component of the magnetic field. In the limit of strong turbulence
numerical simulations suggest that $D(E) \propto E$. Monte-Carlo
simulations and analytical approximations for diffusive propagation in
different regimes have been studied in Ref.~\cite{Casse:2001be}. The
product $D(E)\tau_{\rm ioniz}$ provides an estimate of the (square of
the) distance traveled by the diffusing particle in the case where
energy losses are negligible. The distance traveled by positrons
before being stopped is 
\begin{equation}
d  =  \sqrt{2 D(E) \tau_{\rm ioniz}} \approx   
50\, {\rm pc} \ \gamma^{\frac{2}{3}}
\left( \frac{B}{1 \mu{\rm G}} \right)^{-\frac{1}{6}}
\left( \frac{l_{\rm cell}}{50 {\rm pc}}  \right)^{\frac{1}{3}} 
\left( \frac{N_H}{10^5 {\rm m}^{-3}} \right)^{-\frac{1}{2}}.
\label{d}
\end{equation}
If the regular component is significant, it can help spread the
positrons quickly over the entire bulge. For example, a straight filament
of magnetic field with a $\sim \mu G$ strength and a negligible turbulent
component can transport an electron by a distance as large as $c\tau_{\rm
ioniz}$ away from the production site.  Magnetic field filaments are known to
exist in the galactic center, although their global configuration throughout
the bulge is not known~\cite{Morris:1996th}. 
Therefore, eq.~(\ref{d}) gives a 
lower bound on the size of the positron cloud from a single GRB. 

{\bf GRB Rate --} We now turn our attention to the estimate of the
number of positrons injected in the Galaxy. Gamma ray bursts produce
photons with energies from about 100 keV to 1 MeV over a relatively
short time scale, 1--100 seconds.  Based on observations of the
afterglows associated with {\it long} GRBs, one concludes that they
have cosmological origin and the energy of each burst should be of
order $10^{52}-10^{53}$~ergs, assuming isotropy. However, there is  
convincing evidence that the gamma-ray emission is strongly beamed,
with a characteristic opening angle $\Omega \sim 10^{-3}$. In the case
of a uniform beam (see below for a discussion of structured jets), the
energy emitted per GRB  is~\cite{Frail:2001qp} 
\begin{equation}
E_\Omega \sim 5 \times 10^{50} \ {\rm erg }. 
\label{energy}
\end{equation}
Although there is some variation in the observed energy and the opening
angle, the combined energy estimate in eq.~(\ref{energy}) is fairly 
robust~\cite{Frail:2001qp,Perna:2003bi}.   

Recent observations have provided a strong evidence that GRB is a
supernova-like event~\cite{sn2003}.  Supernovae release an energy
$E_{\rm SN} \sim 10^{53}$~erg, 99\% of which is carried away by
neutrinos, while the remaining 1\% can accommodate the energetic
requirements of a GRB (assuming some reasonable beaming). 

The short time variability on the scale $\delta t \sim 10$~msec of
observed gamma-ray signals suggests that the size of the emission
region is small, $R<c \delta t \approx 3000$~km.  However, the
generation of such a large energy in a small region of space would
result in massive energy losses due to $e^{+}e^{-}$ pair
production. The corresponding optical depth would be as large as $\tau
\sim 10^{11}$, which is, of course, unacceptable. To ameliorate this
problem, one assumes that the emission occurs from a relativistically
moving region that has a large Lorentz factor,
$10^{2-3}$~\cite{Piran:1999kx,Zhang:2003uk}. In such a model the
optical depth can be reduced to $\tau<1$, so that photons can escape
from the emission region. In addition, $\tau<1$ is consistent with the
observed {\em non-thermal} spectrum of photons. 

The observed {\it local} GRB rate is~\cite{Perna:2003bi,Schmidt2001}
\begin{equation}
R^{\rm local}_{\rm observed} = 0.5 \, {\rm Gpc}^{-3} {\rm yr}^{-1}.
\end{equation}
Taking into account the beaming effect, one infers the actual rate of
GRB, which is a factor $(4\pi/\Omega) \sim 500 f_\Omega$ higher.  In
the simplest jet models $f_\Omega$ is a parameter of order
one~\cite{Frail:2001qp}. Therefore, for the total observed rate we get
\begin{equation}
R^{\rm local}_{\rm actual} = 250 \, f_\Omega \, {\rm Gpc}^{-3} {\rm
yr}^{-1}, \ \  f_\Omega \sim 1 .
\end{equation}
Assuming a local density of galaxies $n_{G} =0.01 \, {\rm Mpc}^{-3}$,
we find the GRB rate in a galaxy like our own to be 
\begin{equation}
R^{\rm galaxy}_{\rm actual}  = 25 f_\Omega  \ 10^{-6} {\rm yr}^{-1}.
\label{rate_per_galaxy}
\end{equation}
Of course, this estimate assumes that our galaxy is ``average'' in
terms of the GRB rate. This assumption may be wrong in two ways.  On
the one hand, the GRB rate may be related to the number of stars, in
which case the actual rate is higher than our estimate in
eq.~(\ref{rate_per_galaxy}) because our galaxy is bigger than average.
On the other hand, the rate may be lower if for some reason GRBs occur
predominantly in smaller galaxies or in galaxies morphologically
different from the Milky Way. The afterglows are more frequently
detected in smaller, blue galaxies~\cite{LeFloc'h:2003yp}, but, as
emphasized in Ref.~\cite{LeFloc'h:2003yp}, this may be due to a
selection effect: one is more likely to detect an optical afterglow
from an unobscured galaxy than from a galaxy with a large amount of
dust.  Le Floc'h {\it et al.}~\cite{LeFloc'h:2003yp} argue that the
optically dark GRBs may originate from dust-enshrouded regions of star
formation. Eq.~(\ref{rate_per_galaxy}) implies a galactic GRB output
in $\gamma$-rays
\begin{equation}
E_{{\rm tot}, \gamma} \sim (5 \times 10^{50}) \, 25 \,
  f_\Omega\frac{\rm erg}{(10^6   \, {\rm yr})  {\rm (galaxy)}} 
\sim 10^{52}\,   \frac{\rm MeV} {{\rm (yr)}  {\rm
(galaxy)}} .
\label{Etot}
\end{equation}
This estimate was derived under the simplifying assumption of
``uniform'' beaming, i.e. the GRB jet is assumed to be uniform across
the entire opening angle. Jets could also be ``structured'', with
higher Lorentz factors near the rotation axis (see
e.g. Ref.~\cite{Rossi:2001pk}). In this case, the energy per GRB and
the ratio of the observed to true rate, receive opposite corrections,
leaving the total energy output of GRBs
unchanged~\cite{Perna:2003bi}. Eq.~(\ref{Etot}) thus represents a
reasonable estimate also for jet structures more complicated than a
simple uniform beam. 

{\bf Positrons injection rate --} We write the total energy in
electrons and positrons as 
\begin{equation}
E_{{\rm tot},{(e^+e^-)}} = \zeta_{(e^+e^-)}  E_{{\rm tot}, \gamma}
\label{e_tot}
\end{equation}
where $\zeta_{(e^+e^-)}$ is the ratio to be determined. {\em A priori},
the value of $\zeta_{(e^+e^-)}$ is not known and, in a supernova, a
``natural'' value would be much greater than one. However, a successful
model of GRB must explain the observed non-thermal spectrum of photons.
This implies that the photons are emitted from a fireball which is
sufficiently transparent~\cite{Piran:1999kx,Zhang:2003uk}. Note,
however, that even for $\tau \le 1$, $e^{+}e^{-}$ pairs can be
produced by different mechanisms with efficiencies up to
$\zeta_{(e^+e^-)} \sim O(1)$.

The most efficient mechanism of positron production is pair production
of GRB photons in the field of nuclei and electrons of the
interstellar medium, and we denote by $\zeta_{\rm pair}$ the fraction
of the GRB energy transferred to these positrons. The environment at
the Galactic center is very different from e.g. the environment in the
spiral arms. Dense molecular clouds populate the so-called Central
Molecular Zone (CMZ) which is located in the center of the Galaxy and
which is about 400~pc in diameter~\cite{Morris:1996th}. The
distribution of molecular gas in the CMZ is far from uniform, with
clumps of density $10^5 \,{\rm cm}^{-3}$, embedded in a lower density
$\sim 10^{3.7} \,{\rm cm}^{-3}$ medium. Even higher densities have
been argued necessary, typically $\sim 10^{4} \ (75 {\rm pc} / R_{\rm
  gc})^{1.8} \,{\rm cm}^{-3}$, in order for the molecular clouds at
distance $R_{\rm gc}$ from the Galactic center to survive tidal
stresses~\cite{Gusten:2004ry}.

Gamma-rays from GRB explosions crossing the CMZ produce pairs in the
field of nuclei and electrons of the CMZ gas, with optical depth
$\tau_{_{\rm CMZ}} = \sigma_{pp} n_{_{\rm CMZ}} l_{_{\rm CMZ}}$ where
$\sigma_{pp}$ is the cross section for pair production in the fields
of atoms, both electron and nuclear fields, while $l_{CMZ}$ and
$n_{CMZ}$ are respectively the size and the density of the CMZ. For a
photon of 10 MeV, we find for Hydrogen $\sigma_{pp}^{\rm H} \sim 0.3
\times  10^{-2}$ barn/atom, and for Helium $\sigma_{pp}^{\rm He} \sim
1.1 \times 10^{-2}$ barn/atom, while heavier elements have even larger
cross-sections~\cite{Eidelman:2004wy}. We take $\sigma_{pp} \sim 0.5
\times 10^{-2}$~barn/atom as a conservative value of the cross-section
for the mixture of Hydrogen and heavier elements in the CMZ. Adopting a
fiducial gas density of $10^4 \,{\rm cm}^{-3}$ (note that the presence
of clumps may imply a much larger optical depth) and $l_{CMZ}=400$ pc,
we find an optical depth $\tau \sim \zeta_{\rm pair} \sim 0.03$.

Thanks to the strong, correlated magnetic fields~\cite{Morris:1996th}
in the central region, positrons produced in CMZ would be ejected into
the bulge on timescales comparable to the crossing time, which is
smaller than the (ionization) energy loss time even at these high
densities. So most of the positrons can escape from the CMZ into the
bulge.  Let us denote $\zeta_{_{\rm CMZ}}$ the fraction of the GRB
that either occur inside CMZ or just outside CMZ, with the jet going
through the CMZ. Based on the stellar density in the bulge, as well as
the observations of the afterglows, one can take $\zeta_{_{\rm
    CMZ}}\sim 0.1$, also in view of the fact that despite its small
size, the Galactic bulge contains around 10\% of all the Galactic
molecular Hydrogen and young stars in the Galaxy.

The spectrum of positrons depends on the spectrum of photons escaping
from the fireball. Even assuming that the average energy 
of GRB photons is above 1 MeV, we estimate the production rate of 
these positrons to be:
\begin{equation} 
\Gamma_{(e^+e^-), {\rm cr}} \sim
\zeta_{\rm pair} \, \zeta_{_{\rm CMZ}} \, 
\frac{E_{{\rm tot},{(e^+e^-)}}}{E_{(e^+)}} 
\sim 
\left ( \frac{\zeta_{\rm pair} }{0.03} \right )
\left ( \frac{\zeta_{_{\rm CMZ}}}{0.1} \right )
\left ( \frac{5 \ {\rm MeV}}{E_{(e^+)}} \right )
10^{49} {\rm yr}^{-1} . 
\label{Gamma_cr}
\end{equation}
The injection rate of positrons then appears to be somewhat lower
than the annihilation rate required by eq.~(\ref{annihilation_rate}).
The present scenario thus has to face the same astrophysical
uncertainties that make the intepretation in terms of e.g. type Ic
Supernovae, problematic. We stress, however, that although the naive
estimate of  the optical depth adopted above makes the scenario
marginally compatible with observations, it may underestimate the
actual optical depth, which is sensitive not only to the average
density, but also to the actual distribution of gas. Another possible
way of increasing $\Gamma_{(e^+e^-), {\rm cr}}$ is to assume a
higher-than-average local GRB rate, possibly associated with an
intense starburst in our Galaxy, as suggested by the presence of a
large-scale bipolar wind at the Galactic
center~\cite{Bland-Hawthorn:2002ij}.

Are there other sources of positrons in this scenario? First, some
positrons can be pair-produced outside the fireball by the outgoing
photons interacting with the interstellar medium. Observations support
this. For example, GRB 940217 was observed simultaneously by BATSE,
COMPTEL, and EGRET in the energy range from 0.3~MeV to 100 MeV.  The
spectrum exhibits a significant break at around 1~MeV~\cite{comptel}.
The break point, $1$~MeV, is right at the threshold of pair
production, and is likely to be caused by the $e^{+}e^{-}$ production,
as suggested by Winkler {\em et al.}~\cite{comptel}. Furthermore,
photons can produce pairs by interacting with the fireball photons
which are back-scattered by the
medium~\cite{Beloborodov:2001nz}. However, it has been shown that
typically the  number of positrons produced by this mechanism is
rather small~\cite{Beloborodov:2001nz}. Another source is positron 
leakage directly out of the fireball. The photon pressure inside the
fireball, at temperatures $\sim $keV, is much greater than the
magnetic pressure~\cite{Pe'er:2003ft}, and the mean free path and
gamma-factor of positrons is comparable to those of photons. Finally,
some fraction of the positrons can stay inside the fireball and escape
much later, with a greatly reduced gamma-factor. These positrons stay
close to the site of the GRB and can be used to identify the location
of a recent nearby GRB~\cite{Furlanetto:2002sb}. All these mechanisms
do not appear to increase significantly the number of positrons
predicted by eq.~(\ref{Gamma_cr}), in other words $\zeta_{(e^+e^-)} \sim
\zeta_{\rm pair}$.

{\bf Discussion --} Given the uncertainties associated with this and
other scenarios, it is important to search for signatures that can
help distinguishing GRBs from other sources of positrons. First, one
can test the model by looking for the 511~keV line emission from
nearby galaxies, in particular the nearby Draco and Sagittarius dwarf 
galaxies.  The number of stars, the star formation
rates~\cite{Hodge:1989dj}, and, hence, the rate of the GRBs are much
lower in these dwarf galaxies than in the bulge of the Milky Way. The
expected flux from the dwarfs is suppressed by many orders of
magnitude due to both the low GRB rate and the larger distance.  

Boehm {\em et al.}~\cite{Boehm:2003bt} have estimated the emission
from the nearby galaxies in the case of annihilating dark matter,
which depends on the integral of the square of dark matter
density. They concluded that, since the dwarf galaxies are dark matter
dominated, a large flux could be expected, in contrast with the GRB
case. The observation of the 511~keV emission from the dwarf galaxies
would favor the dark matter scenario.  In Ref.~\cite{Cordier:2004hf}
the search for a 511~keV line from the Sagittarius Dwarf Galaxy (SDG)
was reported, and no such emission was detected. However, the
effective observation time of the SDG was only 80~ks, and the limits
inferred from the data cannot rule out the MeV dark matter as the
origin of the galactic positrons. In the GRB scenario, the large
nearby galaxies, such as M31, could be easier to observe than the
dwarf galaxies, although the flux from M31 would still be suppressed
by a factor $\sim 10^4$ with respect to the Galactic bulge due to the
larger distance.

Integral/SPI observations~\cite{Knodlseder:2005yq} show the ratio of
the contributions from the bulge (B) and from the disk (D) in the
range $3< (B/D)<9$. This is consistent with what predicted from our
scenario, since the molecular clouds in the disk are less dense, and
the production of positron-electron pairs is less efficient. We stress
that an eventual observation with higher spatial resolution (e.g. with
the IBIS instrument, also on INTEGRAL), would clarify whether the 511
KeV emission is actually diffuse, or due to a limited number of point
sources. In the latter case, all scenarios involving a diffuse source
of positrons (as in the case of light Dark Matter), or those involving
several sources in a small region at the Galactic center, with
positrons subsequently spreading in the bulge due to the presence of
regular magnetic fields (as the one presented here), would be
essentially ruled out. The intepretation in terms of production in or
around objects belonging to an old stellar population would instead be
favoured.

If supernovae of a new type, SNe Ic, are associated with GRBs, the
positrons are produced both through the pair production $\gamma \gamma
\rightarrow e^+e^-$ during the burst and by decaying
nuclei~\cite{Casse:2003fh}.  However, the former mechanism dominates
over the latter in each supernova that also produces a GRB. This is
reflected in the rates: we need one GRB per million years to explain
the same population of positrons that could be produced by 200 SNe Ic
per million years, as estimated in Ref.~\cite{Casse:2003fh}. If only a
fraction of supernovae produce GRBs, the amount of pair-produced
positrons in the galaxy exceeds that from decaying nuclei as long as
the rate of GRB per supernova is greater than $5 \times 10^{-3}$.

To summarize, galactic gamma-ray bursts could be responsible for the
population of positrons inferred by the observed 511~keV line from the
galactic center, but only for higher-than-average values of the local
GRB rate or of the values of the optical depth for pair production (in
the field of molecular clouds nuclei) higher than what the estimate
presented here would suggest. Although such values are not
unreasonable, especially in view of the large uncertainties associated
with both quantities, the scenario appears to suffer of the same
difficulties of other astrophysical scenarios discussed in the
introduction and should probably be regarded as only marginally
consistent with observations.

An alternative solution was recently proposed in
Ref.~\cite{Parizot:2004ph}, where the authors suggested that the
dynamics of GRB fireballs could be modified in the innermost regions
of the Galaxy, since remnants of GRBs occurring in the hot,
low-density  medium produced by recurrent starbursts in the CMZ,
become subsonic before they can form a radiative shell. In the
formalism introduced above, this scenario would translate into a very
efficent conversion of the total GRB energy output into  $e^+e^-$
pairs, i.e. $\zeta_{(e^+e^-)} \sim O(1)$. Future observations should
be able to distinguish between this and other proposed models.  

The authors thank J.~Beacom and M.~Kaplinghat for useful discussions.
G.~B. thanks the TEP group at UCLA for hospitality. S.~P.-R. and
S.~P. thank the Theoretical Astrophysics group at Fermilab for
hospitality.  This work was supported in part by the DOE grant 
DE-FG03-91ER40662 and the NASA ATP grants NAG~5-10842 and NAG~5-13399.

\end{document}